\documentclass[aps,prl,preprint,tightenlines,superscriptaddress,showpacs,byrevtex]{revtex4}
\usepackage{graphicx}
\usepackage{dcolumn}
\usepackage{epsfig} 

\newcommand{\dpl}{$D^{+}$}

\newcommand{\RM}{\ensuremath{M_{\mathrm{recoil}}}}
\newcommand{\RMD}{\ensuremath{\Delta M_{\mathrm{recoil}}}}

\newcommand{\dstp}{$D^{*+}$}

\newcommand{\dstm}{$D^{*-}$}
\newcommand{\dstb}{$D^{(*)+}$}

\newcommand{\eetodd}{$e^{+}e^{-} \rightarrow D^{+}D^{-}$}

\newcommand{\eetoddst}{$e^{+}e^{-} \rightarrow D^{+}D^{*-}$}
\newcommand{\eetodstdst}{$e^{+}e^{-} \rightarrow D^{*+}D^{*-}$}
\newcommand{\eetodstdstb}{$e^{+}e^{-} \rightarrow D^{(*)+}D^{(*)-}$}

\graphicspath{{ps}}
\begin{document}
\affiliation{Aomori University, Aomori}
\affiliation{Budker Institute of Nuclear Physics, Novosibirsk}
\affiliation{Chiba University, Chiba}
\affiliation{Chuo University, Tokyo}
\affiliation{University of Cincinnati, Cincinnati, Ohio 45221}
\affiliation{University of Frankfurt, Frankfurt}
\affiliation{Gyeongsang National University, Chinju}
\affiliation{University of Hawaii, Honolulu, Hawaii 96822}
\affiliation{High Energy Accelerator Research Organization (KEK), Tsukuba}
\affiliation{Hiroshima Institute of Technology, Hiroshima}
\affiliation{Institute of High Energy Physics, Chinese Academy of Sciences, Beijing}
\affiliation{Institute of High Energy Physics, Vienna}
\affiliation{Institute for Theoretical and Experimental Physics, Moscow}
\affiliation{J. Stefan Institute, Ljubljana}
\affiliation{Kanagawa University, Yokohama}
\affiliation{Korea University, Seoul}
\affiliation{Kyoto University, Kyoto}
\affiliation{Kyungpook National University, Taegu}
\affiliation{Institut de Physique des Hautes \'Energies, Universit\'e de Lausanne, Lausanne}
\affiliation{University of Ljubljana, Ljubljana}
\affiliation{University of Maribor, Maribor}
\affiliation{University of Melbourne, Victoria}
\affiliation{Nagoya University, Nagoya}
\affiliation{Nara Women's University, Nara}
\affiliation{National Kaohsiung Normal University, Kaohsiung}
\affiliation{National Lien-Ho Institute of Technology, Miao Li}
\affiliation{Department of Physics, National Taiwan University, Taipei}
\affiliation{H. Niewodniczanski Institute of Nuclear Physics, Krakow}
\affiliation{Nihon Dental College, Niigata}
\affiliation{Niigata University, Niigata}
\affiliation{Osaka City University, Osaka}
\affiliation{Osaka University, Osaka}
\affiliation{Panjab University, Chandigarh}
\affiliation{Peking University, Beijing}
\affiliation{Princeton University, Princeton, New Jersey 08545}
\affiliation{RIKEN BNL Research Center, Upton, New York 11973}
\affiliation{Saga University, Saga}
\affiliation{University of Science and Technology of China, Hefei}
\affiliation{Seoul National University, Seoul}
\affiliation{Sungkyunkwan University, Suwon}
\affiliation{University of Sydney, Sydney NSW}
\affiliation{Tata Institute of Fundamental Research, Bombay}
\affiliation{Toho University, Funabashi}
\affiliation{Tohoku Gakuin University, Tagajo}
\affiliation{Tohoku University, Sendai}
\affiliation{Department of Physics, University of Tokyo, Tokyo}
\affiliation{Tokyo Institute of Technology, Tokyo}
\affiliation{Tokyo Metropolitan University, Tokyo}
\affiliation{Tokyo University of Agriculture and Technology, Tokyo}
\affiliation{Toyama National College of Maritime Technology, Toyama}
\affiliation{University of Tsukuba, Tsukuba}
\affiliation{Utkal University, Bhubaneswer}
\affiliation{Virginia Polytechnic Institute and State University, Blacksburg, Virginia 24061}
\affiliation{Yokkaichi University, Yokkaichi}
\affiliation{Yonsei University, Seoul}
  \author{K.~Abe}\affiliation{High Energy Accelerator Research Organization (KEK), Tsukuba} 
  \author{K.~Abe}\affiliation{Tohoku Gakuin University, Tagajo} 
  \author{N.~Abe}\affiliation{Tokyo Institute of Technology, Tokyo} 
  \author{R.~Abe}\affiliation{Niigata University, Niigata} 
  \author{T.~Abe}\affiliation{High Energy Accelerator Research Organization (KEK), Tsukuba} 
  \author{I.~Adachi}\affiliation{High Energy Accelerator Research Organization (KEK), Tsukuba} 
  \author{Byoung~Sup~Ahn}\affiliation{Korea University, Seoul} 
  \author{H.~Aihara}\affiliation{Department of Physics, University of Tokyo, Tokyo} 
  \author{M.~Akatsu}\affiliation{Nagoya University, Nagoya} 
  \author{M.~Asai}\affiliation{Hiroshima Institute of Technology, Hiroshima} 
% ...
\collaboration{The Belle Collaboration}

\preprint{\vbox{ \hbox{   }
                 \hbox{BELLE-CONF-0332}
                 \hbox{EPS Parallel Session: 10}
                 \hbox{EPS-ID 556}
}}

\title{ \quad\\[0.5cm] \Large Measurement of the \eetodstdstb cross-sections}


\noaffiliation

\begin{abstract}
In this paper we report the first measurement of \eetodstdstb\
processes. The cross-sections for \eetodstdst\ and \eetoddst\ at
$\sqrt{s}=10.58\mathrm{GeV}/c^2$ have been measured to be $0.65 \pm
0.04 \pm 0.07\, \mathrm{pb}$ and $0.71 \pm 0.05 \pm 0.09\,
\mathrm{pb}$, respectively. We set an upper limit on the cross-section
of \eetodd\ of $0.04\, \mathrm{pb}$ at the $90\%$ confidence level. In
addition we have measured the fraction of the $D^{*\pm}_T D^{*\mp}_L$
final state in the \eetodstdst\ reaction to be $(97\pm 5)\%$.  The
analysis is performed using $88.9\,\mathrm{fb}^{-1}$ of data collected
by the Belle detector at the $e^{+}e^{-}$ asymmetric collider KEKB.

\end{abstract}

\maketitle
\tighten

\noindent
The processes $e^{+}e^{-} \to D^{(*)} \overline{D} \hspace*{-0.12cm}
~^{(*)}$, with no extra fragmentation particles in the final state,
have not previously been observed at energies $\sqrt{s} \gg 2M_{D}$.
The cross-sections for these processes can be derived once the charmed
meson form factors are determined for the appropriate value of
momentum transfer, $q^2 \equiv s$. In the HQET approach based on the
heavy-quark spin symmetry, the heavy meson form factors are
represented in terms of a universal form factor, called the Isgur-Wise
function.  However, for asymptotically large $q^2 \gg M_c^3/\Lambda$
the leading-twist contribution, which violates the heavy-quark spin
symmetry, becomes dominant~\cite{neubert}.  For an intermediate range
of momentum transfer, sub-leading twist corrections are also
important.  A calculation that takes these effects into
account~\cite{neubert} predicts cross-sections of about $5\,
\mathrm{pb}^{-1}$ for $e^{+}e^{-} \to D
\overline{D}\hspace*{-0.1cm}~^*$ and $e^{+}e^{-} \to D^*_T
\overline{D}\hspace*{-0.1cm}~^*_L$ at $\sqrt{s} \sim 10.6\,
\mathrm{GeV}$ (the subscripts indicate transverse [T] and longitudinal
[L] polarization of the $D^*$); the cross-section for $e^{+}e^{-} \to
D \overline{D}$ is expected to be suppressed by a factor of $\sim
10^{-3}$.

In this paper, we present the first observation of the high energy
\eetodstdst\ and \eetoddst\ processes, and a measurement of their
cross-sections and polarizations.  We also set an upper limit on the
cross-section for \eetodd.

The present study is limited to final states that contain charged
$D^{(*)}$ mesons only: this simplifies the analysis, since it is
possible to select events with no neutral particles in the final
state.  Since the contribution of electromagnetic current coupled to
light quarks is negligible compared to that for heavy quarks, the
neutral and charged charm cross-sections are expected to be the same
\cite{neubert}.


The analysis is based on $88.9\,\mathrm{ fb}^{-1}$ of data at the
$\Upsilon(4S)$ resonance and nearby continuum, collected with the
Belle detector~\cite{Belle} at the KEKB asymmetric energy storage
rings~\cite{KEKB}.  We select well-reconstructed tracks consistent
with originating from the interaction region as charged pion
candidates. Those passing particle identification cuts based on
$dE/dx$, aerogel \v{C}erenkov, and time-of-flight
information~\cite{Belle} are selected as charged kaon candidates.  We
then reconstruct $D^0$ and $D^+$ mesons in the decay modes $D^0 \to
K^- \pi^+$, $D^0 \to K^- \pi^+ \pi^+ \pi^-$ and $D^+ \to K^- \pi^+
\pi^+$. The selected combinations are constrained to a common
vertex, and quality cuts are imposed on the vertex fit to reduce the
combinatorial background. A $15\,\mathrm{MeV}/c^2$ interval around
the nominal $D$ masses is used to select $D^0 \to K^- \pi^+$ and $D^+
\to K^- \pi^+ \pi^+$ candidates; for the $D^0 \to K^- \pi^+ \pi^+
\pi^-$ decay mode the signal window is chosen to be $
10\,\mathrm{MeV}/c^2$ around the nominal $D^0$ mass. The selected $D$
candidates are then subjected to a mass and vertex constrained  fit to
improve their momentum resolution.  The $D^{*+}$ mesons are
reconstructed in the $D^0 \pi^+$ decay mode. The invariant mass of the
$D^0 \pi^+$ combination is required to be within a $
2\,\mathrm{MeV}/c^2$ ($\sim 3 \,\sigma$) mass interval around the
nominal $D^{*+}$ mass.

The processes $e^+ e^- \to D^{(*)+} D^{(*)-}$ can be identified by
energy-momentum balance in fully reconstructed events that contain
only a pair of charm mesons.  However, the small charm meson
reconstruction efficiency via the studied channels results in a very
small total efficiency in this case. Taking into account two body
kinematics, it is sufficient to reconstruct only one of the two
charmed mesons in the event to identify the processes of interest. For
simplicity, we refer to the fully reconstructed $D$ meson as the
\dstb, and the other as the \dstm; the charge-conjugate modes are
included in the analysis. We choose the mass of the system recoiling
against the reconstructed \dstb\ ($\RM(D^{(*)+})$) as a discriminating
variable:
\begin{equation}
\RM(D^{(*)+})=\sqrt{(\sqrt{s}-E_{D^{(*)+}})^2-\vec{p}_{D^{(*)+}}^{~2}},
\end{equation}
where $\sqrt{s}$ is the total center of mass (CMS) energy, and
$E_{D^{(*)+}}$ and $\vec{p}_{D^{(*)+}}$ are the CMS energy and
momentum of the reconstructed $D^{(*)+}$. For the signal a 
peak in the \RM\ distribution around the nominal mass of the recoiling
$D^-$ or $D^{*-}$ is expected. This method provides a significantly
increased efficiency, but also a higher background, in comparison to
full event reconstruction. For the $e^+ e^- \to D^+ D^{*-}$ and $e^+
e^- \to D^{*+} D^{*-}$ processes we find a better compromise between
higher statistics and smaller background: the first $D^{(*)+}$ is
fully reconstructed, while the recoiling $D^{*-}$ is required to decay
into $\bar{D}^0 \pi^-_{slow}$. The reconstructed $\pi^-_{slow}$
provides extra information that allows us to reduce the background to
a negligible level as explained below.

We calculate the difference between the masses of the systems
recoiling against the $D^{(*)+} \pi^-_{slow}$ combination, and against
the $D^{(*)+}$ alone,
\[
  \RMD\equiv \RM(D^{(*)+} \pi^-_{slow})- \RM(D^{(*)+}).
\]
The variable \RMD\ peaks around the nominal $D^{*+}-D^0$ mass
difference with a resolution of $\sigma_{\RMD}\sim
1\,\mathrm{MeV}/c^2$ as found by Monte Carlo simulation.  For
\eetodstdst and \eetoddst we combine \dstp\ and \dpl\ candidates
together with $\pi^-_{slow}$ and require \RMD\ to be within a $\pm
2\,\mathrm{MeV}/c^2$ interval around the nominal $M_{D^{*+}} -M_{D^0}$
mass difference.

The $\RM(D^{*+})$ and $\RM(D^+)$ distributions are
shown in Figs.~1a and 1b, respectively. Clear signals are
seen around the nominal $D^{*-}$ mass in both cases. The higher recoil
mass tails in the signal distribution are due to initial state
radiation (ISR). The hatched histograms show the \RM\ distributions
for events in the \RMD\ sidebands.
\begin{figure}[htb]
\begin{center}
\hspace*{-0.1cm}\epsfig{file=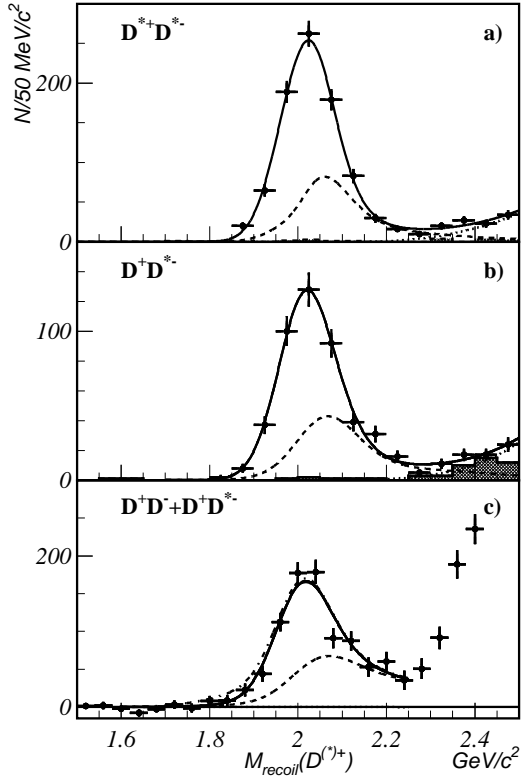,width=0.42\textwidth}
\end{center}
\caption{a-b) The distributions of the mass of the system recoiling
against a) $D^{*+}$, and b) $D^+$.  Points with error bars show the
signal \RMD\ region; hatched histograms correspond to \RMD\
sidebands. The solid lines represent the fits described in the text;
the dashed lines show the contribution due to events with ISR photons
of significant energy.  The dotted lines show the expected background
contribution.  c) The distributions of $\RM(D^+)$ without any
requirement on \RMD. }
\label{dstdst}
\end{figure}

The backgrounds in the region of $\RM < 2.1 \, \mathrm{GeV}/c^2$, are
negligible for both processes. To provide a numerical estimate, we
divide the background sources into three categories:
\begin{itemize}
\item [I] fake reconstructed $D^{*+}$ or $D^+$;
\item [II] $e^+ e^- \to D^{(*)+} D^{(*)} n \pi $, where the
$\pi^-_{slow}$ is not produced from $D^{*-}$ decays, and, thus, produces
no peak in the \RMD\ distribution;
\item [III] $e^+ e^- \to D^{(*)+} D^{*-} n \pi $, where $n \ge 1$.
\end{itemize}

First we consider the backgrounds for the process \eetodstdst.  To
estimate background (I) we count the entries in the $\RM(D^0\pi^+) <
2.1 \, \mathrm{GeV} / c^2$ interval for $D^0 \pi^+$ combinations taken
from the $D^{*+}$ mass sideband ($2.016 < M_{D^0\pi^+} <2.02\,
\mathrm{GeV}/c^2$). Three events are found in the data, while the
Monte Carlo predicts a contribution of 2.5 events the from
non-Gaussian tails in the $M_{D^{*+}}$ resolution function.  The
signal Monte Carlo is normalized to the number of entries in
$\RM(D^{*+}) < 2.1 \, \mathrm{GeV} / c^2$ region in the data.
Background (II) is estimated using \RMD\ sidebands ($0.150 < \RMD <
0.154 \, \mathrm{MeV}/c^2$). In the region $\RM(D^{*+})
<2.1\,\mathrm{GeV}/c^2$ 8 events are found in the data; according to
Monte Carlo, 4 events are expected from initial state radiation. Thus,
backgrounds (I) and (II) are estimated to be smaller than $3$ and
$9$ events at $90\%~\mathrm{CL}$, respectively. As a cross-check we
also study the wrong-sign $D^{(*)+} \pi^+_{slow}$ combinations: 2
events with wrong-sign $\pi_{slow}^+$'s are found in the data in the
interval $\RM(D^{*+}) < 2.1\, \mathrm{GeV}/c^2$.

The remaining background (III) can result in peaks in both the
$M(D^{*+})$ and \RMD\ distributions, but has a threshold in \RM\
distribution at $M_{D^{*+}} + M_{\pi^0} = 2.15 \, \mathrm{GeV} / c^2$,
which is $\sim 1\sigma$ away from the chosen \RM$(D^{*+})$ interval.
To estimate the residual background (III) contribution we perform a
fit to the \RM$(D^{*+})$ distribution. The signal function is the
result of convolving the generated \RM$(D^{*+})$ distribution with the
detector resolution: the signal function is the  sum of a core
Gaussian and an asymmetric function representing the \RM$(D^{*+})$
shape when the studied process is accompanied by radiative photon(s)
with significant energy.  The \RM$(D^{*+})$ resolution due to detector
smearing and the signal function offset are left as free parameters in
the fit, to test the agreement with the Monte Carlo predictions. The
background (III) distribution is parameterized by a threshold
function, $ \alpha \cdot (\RM(D^{*+})-M(D^{*-})_{PDG}
-M(\pi^0)_{PDG})^{\beta}$, convolved with the detector resolution,
where $\alpha$ and $\beta$ are free parameters. The fit results are
shown in Fig.~\ref{dstdst}{\it a} as the solid line. The dotted line
represents the expected background (III) distribution. The signal
yield is found to be $815\pm 28$ events. The \RM\ resolution
$\sigma=56.1 \pm 2.2 \, \mathrm{MeV}/c^2$ is found to be in excellent
agreement with the Monte Carlo expectation ($56.4\,
\mathrm{MeV}/c^2$), and the shift of the signal peak position in the
data with respect to the Monte Carlo position is found to be
consistent with zero ($0.6 \pm 2.5\,\mathrm{MeV}/c^2$).  The
contribution from background (III) in the $\RM < 2.1\,
\mathrm{GeV}/c^2$ interval is estimated from this fit to be less than
$2$ events.

The backgrounds for the \eetoddst\ process are studied in a similar
way.  Five events are found in the data in the $D^+$ mass sideband region
($20<\left| M_{K^-\pi^+ \pi^+}-M_{D^+} \right| < 35 \, \mathrm{MeV} /
c^2$), whereas 1.5 events are expected from the signal Monte Carlo.
There are 6 data events in the \RMD\ sidebands with $\RM(D^+) < 2.1\,
\mathrm{GeV}/c^2$, while 6 signal events are expected from the Monte
Carlo simulation. Thus backgrounds (I) and (II) for the \eetoddst\
process are estimated to be smaller than $7$ and $4$ events at
$90\%~\mathrm{CL}$, respectively. Three wrong sign $D^+ \pi_{slow}^+$
combinations are found in the data in the interval $\RM(D^+) < 2.1\,
\mathrm{GeV}/c^2$.  A similar fit is then performed to the \RM($D^+$)
distribution.  The signal yield is found to be $423 \pm 20$
events. Again, excellent agreement between the Monte Carlo and signal
shape parameters is found: $\sigma=58.1 \pm 3.6 \, \mathrm{MeV}/c^2$
($60 \,\mathrm{MeV}/c^2$ is expected from the Monte Carlo);
$M(D^{*-})_{data} -M(D^{*-})_{MC}=-2.1 \pm 3.6\,\mathrm{MeV}/c^2$. We
conclude that background (III) is smaller than $2$ events.

From the above study we estimate the total background in the $\RM <
2.1\, \mathrm{GeV}/c^2$ interval to be smaller than $14$ and $16$
events for the \eetodstdst\ and \eetoddst\ processes, respectively,
which is of order of  $1\%$ of the signal. We assume all events in
the interval $\RM < 2.1\, \mathrm{GeV}/c^2$ are signal and include
the possible background contribution in the systematic error.

\begin{figure}[htb]
\begin{center}
\epsfig{file=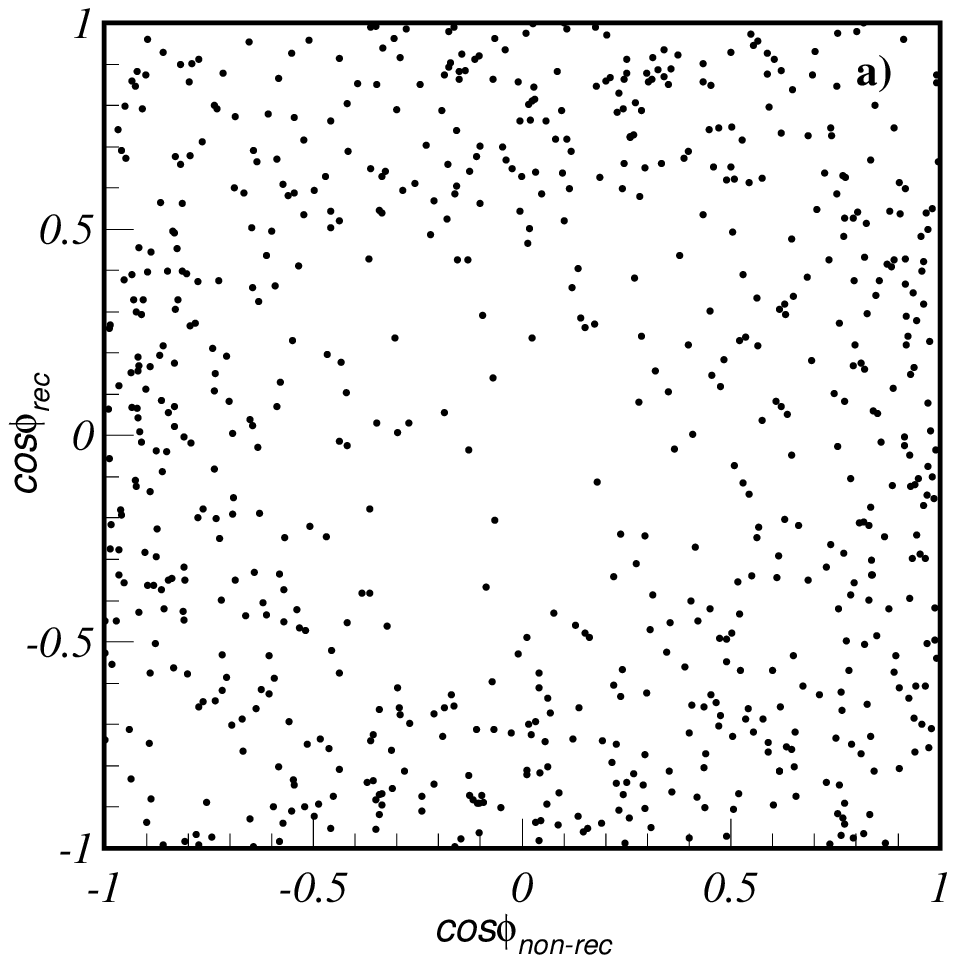, width=0.42\textwidth}
\epsfig{file=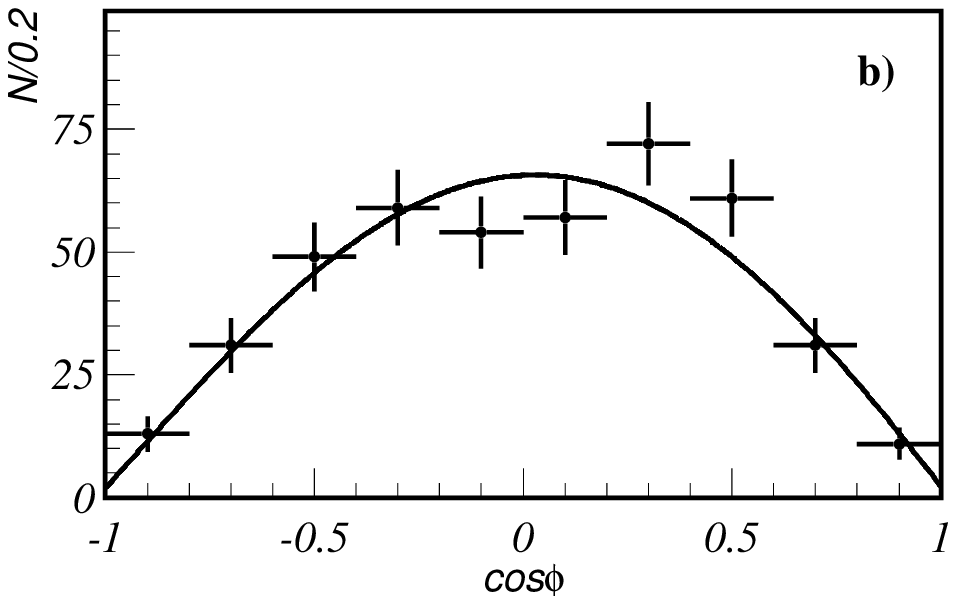, width=0.42\textwidth}
\end{center}
\caption{a) The scatter plot $\cos(\phi_{D^*_{rec}})$ {\emph {vs}}
$\cos(\phi_{D^*_{non-rec}})$ (\eetodstdst). b) $D^{*+}$ meson helicity
angle distribution for (\eetoddst) signal candidates.}
\label{heli}
\end{figure}

Since the reconstruction efficiency depends on the production and
$D^{*\pm}$ helicity angles, we perform an  angular analysis before
computing  cross-sections. The helicity angle of the
non-reconstructed $D^{*-}$ is calculated assuming two-body
kinematics. A scatter plot of the helicity angles for the two
$D^{*}$-mesons from \eetodstdst\ ($\cos\phi(D^*_{rec})$ \emph {vs}
$\cos\phi(D^*_{non-rec})$) for the recoil mass region $\RM(D^{*+})
< 2.1 \, \mathrm{GeV}/c^2$ is shown in Fig.~\ref{heli}{\it a}. The
distribution is fitted by a sum of three functions corresponding to
the $D^*_T D^*_T$, $D^*_T D^*_L$ and $D^*_L D^*_L$ final states,
obtained from Monte Carlo simulation. The fit finds the fractions of
$D^*_T D^*_T$, $D^*_T D^*_L$ and $D^*_L D^*_L$ final states to be
$(1.5\pm 3.6)\%$, $(97.2 \pm 4.8)\%$ and $(1.3 \pm 4.7)\%$,
respectively.  Figure~\ref{heli}{\it b} shows the $D^{*-}$ meson
helicity distribution for \eetoddst. The fraction of the $D^+
D^{*-}_T$ \footnote{There was a typo  in the previous draft version and in the EPS2003 presentation . ``fraction of the $D^+
D^{*-}_L$'' was typed. Now the text is correct.} final state is found from the fit to be equal to $(95.8 \pm
5.6)\%$.

The raw production angle distributions for $D^{*+}$ from \eetodstdst\
and $D^+$ from \eetoddst\ processes are obtained from the region of
recoil masses $\RM < 2.1\, \mathrm{GeV}/c^2$. In this region, the
efficiency for signal events with initial state radiation photon(s)
with significant energy is low. We calculate the fraction of rejected
ISR events according to a Monte Carlo simulation based on
Ref.~\cite{ISR}, in bins of the production angle. The reconstruction
efficiency is estimated from the Monte Carlo simulation, in the same
production angle bins, taking the measured $D^{*\pm}$ polarization
into account.  Figures~3a and 3b show the production angle distributions
for $D^{*+}$ from \eetodstdst\ and $D^+$ from \eetoddst, respectively,
after correcting for the reconstruction efficiencies and intermediate
branching ratios. These distributions are fitted with the function $N
\cdot (1 + \alpha \cos^{2}{\theta})$.  To calculate the total
cross-sections, the signal yields are corrected to take into account
the fraction of events with initial state radiation that lie outside
of the interval $\RM < 2.1\, \mathrm{GeV}/c^2$.  The efficiency
corrected signal yields are found to be $58000\pm 3400$ and $64000\pm
4800$ for the \eetodstdst\ and \eetoddst\ processes respectively. The
parameters $\alpha$ are equal to $0.8 \pm 0.3$ and $2.3_{-0.7}^{+0.8}$
for the two processes.
\begin{figure}[htb]
\begin{center}
\epsfig{file=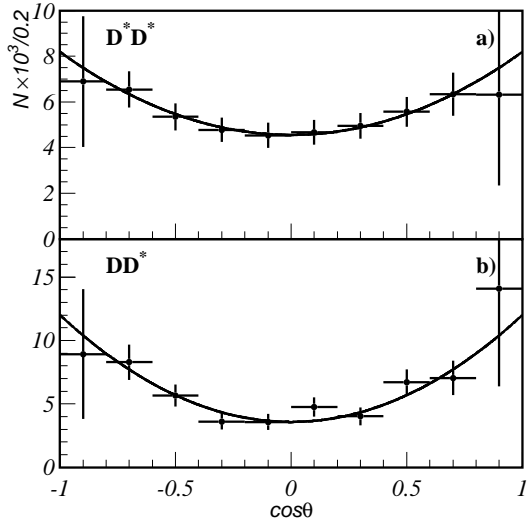,width=0.42\textwidth}
\end{center}
\caption{The $D^{(*)+}$ production angle distribution for a)
\eetodstdst\ and b) \eetoddst\ processes after correcting for
reconstruction efficiencies and intermediate branching fractions.}
\label{results}
\end{figure}

We find  cross-sections of  $0.65 \pm 0.04 \pm 0.07\,
\mathrm{pb}$ and $0.71 \pm 0.05 \pm 0.09\, \mathrm{pb}$ for
\eetodstdst\ and \eetoddst, respectively, where the first error is
statistical and the second systematic.  The sources of systematic
error are summarized in Table~\ref{systematics}. The dominant
contributions are from the uncertainties in tracking efficiency and
corrections for the initial state radiation.
\begin{table}[htb]
\begin{center}
\begin{tabular}{|l|c|c|}
\hline
\ Source  &\ \eetodstdst \ &\ \eetoddst \ \\
\hline
\hline
\ Tracking efficiency \  & 9\%   & 8\%  \\
\ Identification         & 2\%   & 2\%  \\
\ Backgrounds            & $~^{+1}_{-0}\%$   &  $~^{+1}_{-0}\%$ \\
\ ISR correction         & 5\%   & 5\%  \\
\ $\mathcal{B}(D^{(*)})$  & 4\%   & 8\%  \\
\hline
\hline
\ Total                 & 11\%   & 13\% \\
\hline
\end{tabular}
\end{center}
\caption{Contributions to the systematic error for the \eetodstdst\
and \eetoddst\ processes.}
\label{systematics}
\end{table}

We search for the process \eetodd\ by studying the recoiling against
the reconstruted $D^+$ $(M_{recoil})$.  In the \eetodstdst\ and
\eetoddst\ analyses, backgrounds are strongly suppressed by the tight
\RMD\ cut, which is not applicable dor \eetodd. Without this cut the
combinatorial background is significant ($\sim 20\%$). We use $D^+$
mass sidebands ($20< \left| M_{K\pi\pi}-M_D \right| <35 \,
\mathrm{MeV}/c^2 $) to extract the \RM\ distribution for the
combinatorial background. Fig.~1c shows the distribution of \RM$(D^+)$
after sideband subtraction. To extract the \eetodd\ and \eetoddst\
yields we fit this distribution with the sum of two signal functions
corresponding to $D^-$ and $D^{*-}$ peaks and a background function.
The latter is a threshold function, $\alpha \cdot (x-M(D^{-})_{PDG}
-M(\pi^0)_{PDG})^{\beta}$, convolved with the detector resolution,
where $\alpha$ and $\beta$ are free parameters. For the fit we use
only the region $\RM < 2.25\,\mathrm{GeV}/c^2$, because of the
contribution of $e^+ e^- \to D^{(*)} D^{**}$  at higher \RM.

The fit finds $-13\pm 24$ events in the $D^+$ peak and $935\pm 42$ in
the $D^{*+}$ peak.  The fit function is shown in the Fig.~1c as the
solid line; the dashed line shows the contribution of events with ISR
photons of signficant energy (larger in this case due to absence of
the \RMD\ cut); and the dotted line represents the case where the
contribution of \eetodd\ is set at the value corresponding to the
$90\%$ confidence level upper limit.  The reconstruction efficiencies
for the \eetodd\ and \eetoddst\ are found from Monte Carlo to be
$3.1\%$ and $1.7\%$, respectively. In the \eetodd\ Monte Carlo the
$D^+$ production angle is assumed to have a $\sin^2{\theta}$
distribution, as required by conservation of angular momentum; for
\eetoddst\ the production angle distribution is fixed according to the
analysis presented above.  The \eetoddst\ cross-section is calculated
to be $0.61\pm0.05\, \mathrm{pb}$ which agrees with the result using
the \RMD\ method. However, in this case the systematic uncertainty in
the signal yield is large due to the non-negligible $e^+ e^- \to D^+ D
\pi$ background under the peak, which can only be determined from the
higher \RM\ region. For the \eetodd\ cross-section we set an upper
limit of $0.04\,$pb at the $90\%$ confidence level.


In summary, we report the first measurement of the cross-sections for
the \eetodstdst\ and \eetoddst\ processes at $\sqrt{s}=10.6
\mathrm{GeV}$ to be $0.65\pm 0.04\pm 0.07 \, \mathrm{pb}$ and $0.71\pm
0.05\pm0.09 \, \mathrm{pb}$, respectively, and set an upper limit on
the \eetodd\ cross-section of $0.04\, \mathrm{pb}$ at $90\%$
confidence level.  The measured cross-sections are an order of
magnitude lower than those predicted in the Ref.~\cite{neubert}, but
their relative sizes are as predicted: the cross-sections for
\eetodstdst\ and \eetoddst\ are found to be close each other, while
the cross-section for \eetodd\ is much smaller. The helicity
decomposition for \eetodstdst\ is found to be saturated by the
$D^{*\pm}_T D^{*\mp}_L$ final state (the fraction is equal to $(97.2
\pm 4.8) \%$) and for \eetoddst\ -- by the $D^*_T$ final state
($95.8\pm5.6\%$), in good agreement with the predictions of
Ref.~\cite{neubert}.

We wish to thank the KEKB accelerator group for the excellent
operation of the KEKB accelerator.
We acknowledge support from the Ministry of Education,
Culture, Sports, Science, and Technology of Japan
and the Japan Society for the Promotion of Science;
the Australian Research Council
and the Australian Department of Education, Science and Training;
the National Science Foundation of China under contract No.~10175071;
the Department of Science and Technology of India;
the BK21 program of the Ministry of Education of Korea
and the CHEP SRC program of the Korea Science and Engineering Foundation;
the Polish State Committee for Scientific Research
under contract No.~2P03B 01324;
the Ministry of Science and Technology of the Russian Federation;
the Ministry of Education, Science and Sport of the Republic of Slovenia;
the National Science Council and the Ministry of Education of Taiwan;
and the U.S.\ Department of Energy.

\end{document}